\documentclass{cjaa}                   

\usepackage{graphicx}                   
\input{epsf.sty}                        
\input{psfig.sty}                       

\begin{document}

   \title{A Catalog of Luminous Infrared Galaxies in the {\it IRAS} Survey and the Second Data Release of the SDSS
}

   \volnopage{Vol.0 (200x) No.0, 000--000}      
   \setcounter{page}{1}          

   \author{Chen Cao
      \inst{1,2}
      \mailto{}
   \and Hong Wu
      \inst{1}
   \and Jian-Ling Wang
      \inst{1,2}
   \and Cai-Na Hao
      \inst{1,2}
   \and Zu-Gan Deng
      \inst{3,1}
   \and Xiao-Yang Xia
      \inst{4}
   \and Zhen-Long Zou
      \inst{1}
      }
   \offprints{Chen Cao}                   

   \institute{National Astronomical Observatories, Chinese Academy of Sciences,
             Beijing 100012, P. R. China\\
             \email{caochen@bao.ac.cn}
        \and
             Graduate School of the Chinese Academy of Sciences, 100039 Beijing, China\\
        \and
             College of Physical Sciences, Graduate School of the Chinese Academy of Sciences, P.O. Box 3908, 100039 Beijing, China\\ 
        \and
             Department of Physics, Tianjin Normal University, 300074 Tianjin, China\\
          }

   \date{Received~~2005 month day; accepted~~2005~~month day}

   \abstract{
We select the Luminous Infrared Galaxies by cross-correlating the Faint Source Catalogue (FSC) and Point 
Source Catalogue (PSC) of the {\it IRAS} Survey with the Second Data Release of the SDSS for studying 
their infrared and optical properties. The total number of our sample is 1267 for FSC and 427 for PSC by using 
2$\sigma$ significance level cross-section. The "likelihood ratio" method is used to estimate 
the sample's reliability and for a more reliable subsample (908 for FSC and 356 for PSC) selection. 
Then a Catalog with both the infrared, optical and radio informations is presented and will be used in further works. 
Some statistical results show that the Luminous Infrared Galaxies are quite different from the Ultra-Luminous Infrared Galaxies. 
The AGN fractions of galaxies with different infrared luminosities and the radio to infrared correlations are consist with 
previous studies.
   }

   \authorrunning{C. Cao et al.}            
   \titlerunning{A Catalog of LIGs in {\it IRAS} and SDSS-DR2}  

   \maketitle

%
%
\section{Introduction}           
\label{sect:intro}
~~~~The research of Luminous Infrared Galaxies (LIGs, the galaxies with infrared luminosity (L$_{\rm IR}$, 8-1000 $\mu$m) 
higher than 10$^{11}$ L$_{\rm \odot}$) began after the success of the first mid- to far-infrared all-sky survey carried 
out in 1983 by the Infra-Red Astronomical Satellite ({\it IRAS}). The physical properties of the LIGs, especially 
the Ultra-Luminous Infrared Galaxies (ULIGs, L$_{\rm IR}$ $>$ 10$^{12}$ L$_{\rm \odot}$) were studied by using the {\it IRAS} 
infrared data and the follow-up optical (POSS, DSS, {\it HST}, VLT ...) observations, such as the analyses of the 
Bright Galaxy Sample (BGS, Soifer et al. 1987b), the optical spectroscopy of LIGs (Kim et al. 1995; Veilleux et al. 
1995), the statistical study of the spectra of very luminous IRAS galaxies (Wu et al. 1998ab), the IRAS 1 Jy Survey 
of ULIGs (Kim et al. 1998ab) and the Point Source Catalog redshift survey (PSCz, Saunders et al. 2000). From the 
previous studies people found that most of the ULIGs are in an interaction/merger system (Zou et al. 1991; 
Sanders et al. 1988; Kim et al. 1995; Lawrence et al. 1989) and with a high AGN fraction (Kim et al. 1995,2002; 
Wu et al. 1998ab). 
There is a possible evolution path (Sanders et al. 1988; Sanders \& Mirabel 1996) from galaxy mergers to 
quasi-stellar objects (QSOs) and elliptical galaxies, which supports the hierarchical galaxy formation theory 
(Cole et al. 2000). The LIGs with L$_{\rm IR}$ $\sim$ 10$^{11}$ -- 10$^{12}$ L$_{\rm \odot}$ are quite different from the 
ULIGs in their morphologies and spectral features. The recent studies of the distant LIGs (0.4$<$z$<$1.2, Zheng 
et al. 2004) showed that there are many massive disks which have been forming a large fraction of their stellar 
mass since z = 1, and most of their central parts were formed prior to the formation of their disks. Although the 
LIGs are so important for studying, there has not been a large and reliable sample of LIGs for statistical analyses, 
so lots of physical properties the LIGs are still unclear. The role of LIGs and ULIGs in the formation and evolution 
of the galaxies is still a problem to be resolved. 

In order to study the properties of the LIGs in more detail, we need a large sample which has both the infrared and 
optical informations for our analyses. The Sloan Digital Sky Survey (SDSS) was chosen for the cross-correlation 
with {\it IRAS} data because of its large sky coverage ($\sim$2627 deg$^{2}$ for spectroscopic targets of the second 
data release) and high spectral signal-to-noise (S/N) ratio and spectral resolution (R $\sim$ 1800). Although some 
authors have studied the optical properties for {\it IRAS} galaxies using the SDSS data (Goto 2005b; Pasquali et al. 2005), 
their cross-correlation between optical and infrared catalogs is relatively simple (only use a fixed circle) for a 
reliable sample selection and they didn't present a complete catalog for further analyses. 
The structure of this paper is as follows: In Sect.2 we give a simple description of the data and the cross-correlation 
between {\it IRAS} and SDSS; In Sect.3 we use the "likelihood ratio" method for detailed identifications for our 
sample and estimate its reliability; In Sect.4 we describe our Catalog; In Sect.5 we do some 
statistical works based on a selected subsample. Finally the summary is given in Sect.6. We adopt cosmological 
parameters {\it H}$_{\rm 0}$=70 kms$^{-1}$Mpc$^{-1}$, $\Omega$$_{\rm m}$=0.3, $\Omega$$_{\rm \Lambda}$=0.7 throughout this paper.


\section{Data Description and Sample Selection}
\label{sect:Data}
\subsection{{\it IRAS} Faint Source Catalog and Point Source Catalog}
~~~~The Infra-Red Astronomical Satellite ({\it IRAS}) was launched in 1983 (Neugebauer et al. 1984; Soifer et al. 1987a) 
and scanned almost all the sky in mid- and far-infrared (12, 25, 60, 100 $\mu$m) wavebands. The Faint Source Catalog 
(FSC, $\mid$b$\mid$ $>$ 10, Version 2.0, Moshir+ 1989) was released after the Point Source Catalog (PSC, Version 2.0, 
IPAC 1986). It contains data for 173044 point sources in unconfused regions with flux densities typically above 0.2 Jy at 12, 25 
and 60 $\mu$m, and above 1.0 Jy at 100 $\mu$m, achieves roughly one-magnitude deeper in sensitivity relative to the PSC. 
The catalogues (both the FSC and PSC) give the 
{\it IRAS} sources' four band flux densities and qualities, the positions of the sources, and other useful parameters. 
The sources in the catalogues all have large positional uncertainties which can be described as an "error ellipse". 
The error ellipse stands for the uncertainties along (in-scan) and cross (cross-scan) the {\it IRAS}'s scan direction, 
and the uncertainty ellipse major axis, minor axis and positional angle in the catalogues 
are used for describing it. The FSC is deeper than PSC but may be contaminated by foreground and background sources, 
the PSC is shallower but can be used for a comparison with previous results (e.g., the PSCz). Therefore, we use them 
separately to make up our sample and do statistical analyses based on each of them.

\subsection{SDSS-DR2 Data}
~~~~The Sloan Digital Sky Survey (SDSS, York et al. 2000) contains an imaging survey of northern sky in the 
five bands u, g, r, i, z and a spectroscopic target survey performed by multi fibers. The Second Data Release (DR2, 
Abazajian et al. 2004, Version v2$\_$20040928$\_$1505) was released in 2004. The SDSS-DR2 spectroscopic target survey 
covers about 2627 deg$^{2}$ of the sky, including about 260490 galaxies, 32241 quasars, 3791 high-z (z $>$ 2.3) quasars 
and others objects. For the study of the detailed spectral properties of LIGs (such as their emission lines), 
we only choose the SDSS-DR2 spectroscopic targets with the redshift greater than 0.001 (to reject stars) and 
high redshift confidence (zConf $>$ 0.9) to do the cross-correlation. Finally we obtain 268202 sources from SDSS 
datasets as our candidates for the cross-correlation with {\it IRAS} catalogues. 

\subsection{Cross-Correlation between the {\it IRAS} and SDSS}
~~~~We use the {\it IRAS} (FSC and PSC, separately) error ellipse as the cross-section (the SDSS's position 
uncertainties are neglected compared with the {\it IRAS}'s) to do cross-correlation with the SDSS sources 
spectral positions. Two RMS uncertainty (2$\sigma$) significance level was chosen for a high level confidence 
and more complete sample selection. The SDSS spectral redshift and the {\it IRAS} flux densities were then used 
to calculate the infrared luminosity (L$_{\rm IR}$) of the matched sources. Due to the fact that the 12$\mu$m and 25$\mu$m 
flux densities of the objects are mostly the "upper limit" (flux quality = 1), we calculate the far-infrared luminosity 
(Helou et al. 1988; Sanders \& Mirabel 1996) and then convert it to the total infrared luminosity 
(1-1000$\mu$m, Calzetti et al. 2000)$\footnote{The contribution to the total infrared luminosity from the 1-8$\mu$m 
regime is expected to be of the order of a few percent (Calzetti et al. 2000).}$: 
\begin{equation}
F_{\rm FIR} = 1.26 \times 10^{-14} \lbrace 2.58f_{\rm 60}+f_{\rm 100} \rbrace \lbrack Wm^{-2} \rbrack
\label{eq:Lir1}
\end{equation}
\begin{equation}
L_{\rm FIR} = 4 \pi D_{\rm L}^{2} F_{\rm FIR} \lbrack L_{\rm \odot} \rbrack
\label{eq:Lir2}
\end{equation}
\begin{equation}
L_{\rm IR} (1-1000 \mu m) = 1.75 L_{\rm FIR} 
\label{eq:Lir3}
\end{equation}
where f$_{\rm 60}$, f$_{\rm 100}$ are the {\it IRAS} flux densities in Jy at 60 and 100 $\mu$m respectively. 
Then the LIGs (L$_{\rm IR}$ $\geq$ 10$^{11}$ L$_{\rm \odot}$) were chosen as our sample objects and the number 
of sources is 1267 for FSC and 427 for PSC$\footnote{Note that the objects with 60$\mu$m flux quality =1 have 
been rejected. We didn't treat the objects with 100$\mu$m upper limit because it doesn't affect much on the 
calculation of L$_{\rm IR}$ (see Sect 5.2 for details).}$. 
From this sample we present a Catalog (will be described in Sect.4) and 
perform detailed identifications and further analyses. Fig. 1 is the sky coverage of our sample (both FSC and PSC) 
in equatorial coordinates, which shows that it covers nearly all the SDSS-DR2 spectroscopic survey regions.

   \begin{figure}
   \plottwo{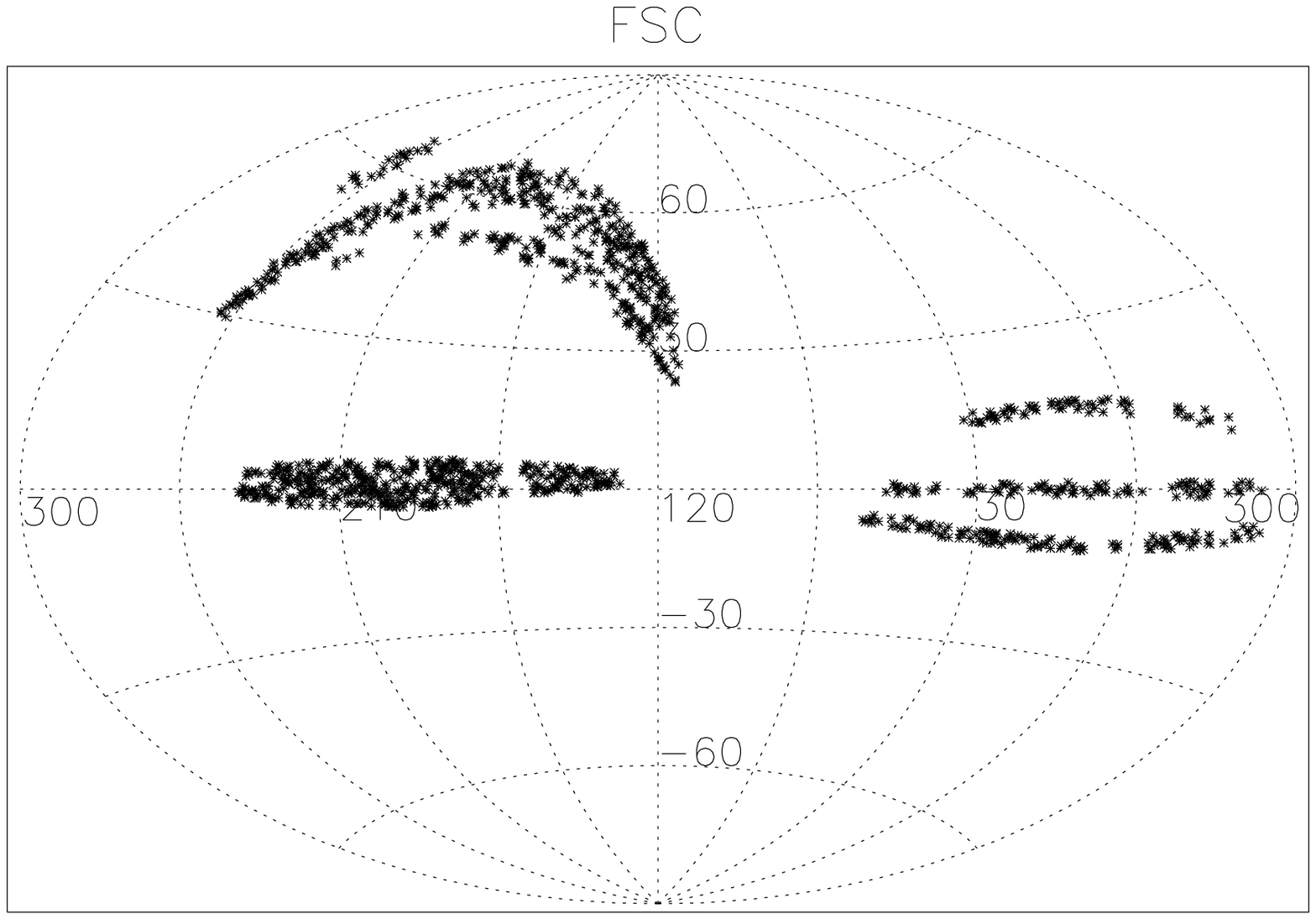} {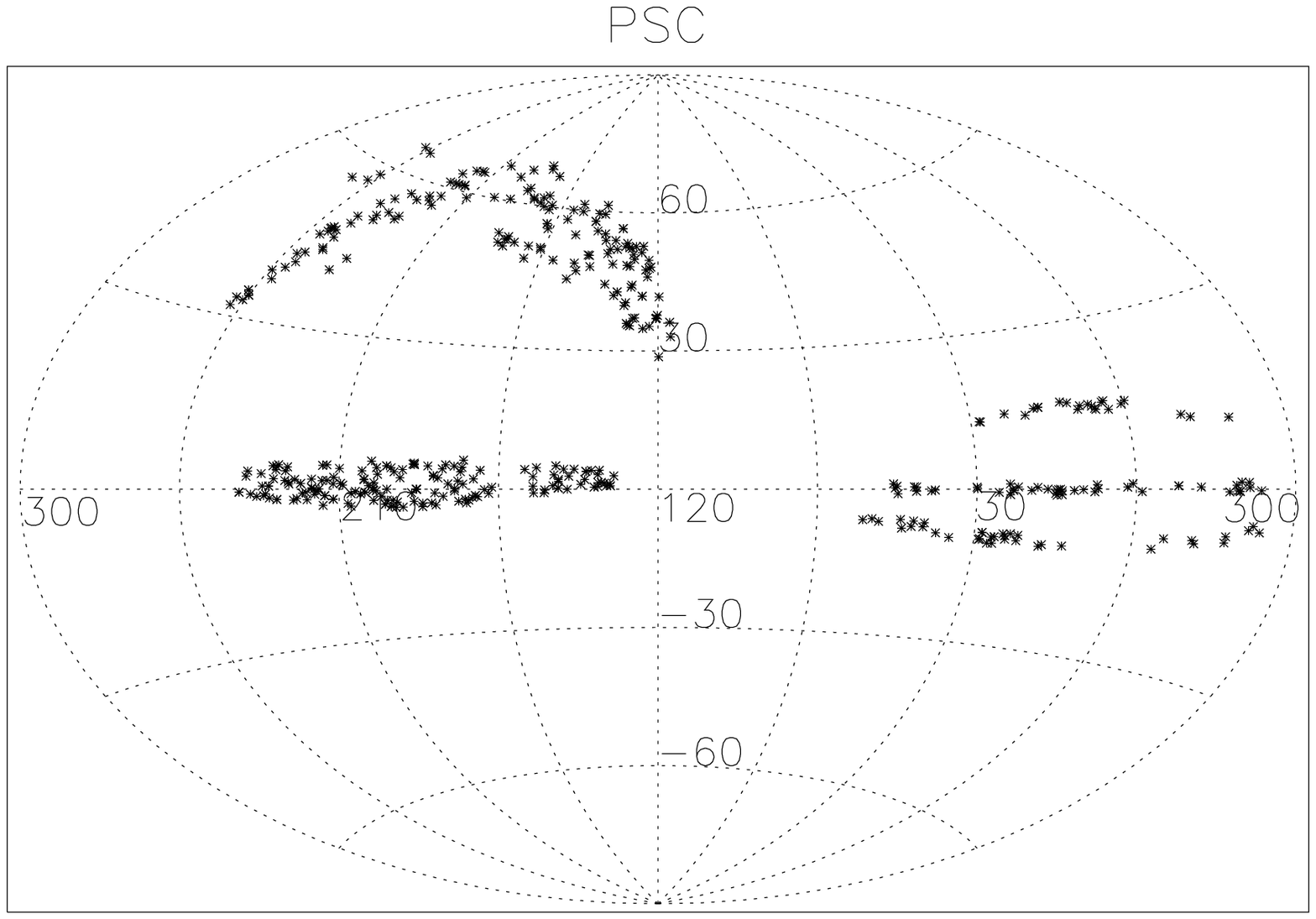}
   \caption{Distribution on the sky of the objects in our sample. This is an Aitoff projection in equatorial coordinates. 
Left: The FSC sample; Right: The PSC sample.}
   \label{Fig:plot1}
\end{figure}

\subsection{VLA-FIRST Data}
~~~~The NRAO Very Large Array (VLA) Faint Images of the Radio Sky at Twenty-centimeters (FIRST) data (Becker et al. 1995) 
are used here for studying the radio properties of our sample. The FIRST survey is a project designed to produce the 
radio equivalent of the Palomar Observatory Sky Survey (POSS) over 10$^{4}$ deg$^{2}$ of the North and South Galactic 
Caps. The FIRST Survey Catalog (White et al. 1997, from the 1993 through 2002, contains $\sim$ 811000 sources and covers 
$\sim$ 9030 deg$^{2}$) including peak and integrated flux densities and size information is generated from the coadded 
images. The individual sources have 90$\%$ confidence error circles of radius $<$ 0.5" at the 3mJy level and 1" at the 
survey threshold ($\sim$ 1mJy). The survey area has been chosen to coincide with that of the SDSS First Data Release 
(DR1) and $\sim$ 50$\%$ of the optical counterparts to FIRST sources will be detected. We use the FIRST Survey Catalog 
updated at 2003 April 11 to perform the cross-correlation with the objects in our sample.

We match our sample's SDSS spectral positions with the VLA FIRST positions using a 2" searching radius and find that 
there are 624 objects for FSC and 258 for PSC which are contained in the FIRST catalog. 
This result means that the radio flux densities of these sources are all above the FIRST's threshold (about 1mJy). 
Thus they have a higher probability to be true IR sources because of the (far-) infrared to radio correlation 
(will be discussed in Sect.5, Helou et al. 1985,1993; Condon 1992; Ivezi\'{c} 2002).

\subsection{Reliability and Completeness}
~~~~Due to our large 2$\sigma$ cross-sections for the cross-correlation, there are also some SDSS objects which 
are not really the IR sources being selected as our sample objects because of the contamination of foreground 
and/or background sources. So we calculate the random probability that the SDSS-DR2 spectroscopic targets fall 
into the {\it IRAS} 2$\sigma$ error ellipse by assuming that the SDSS targets are uniformly distributed across 
the 2627 deg$^{2}$ sky and the mean {\it IRAS} 2$\sigma$ error ellipse area is about 0.56 arcmin$^{2}$ for the LIGs. 
The random probability is about 4.32$\%$ for FSC sample and 5.02$\%$ for PSC and hence our whole sample's reliability 
is about 95.68$\%$ (FSC) and 94.98$\%$ (PSC) (R = 1 - N$_{\rm random}$/N$_{\rm real}$).

The completeness of our sample can be estimated from the 2$\sigma$ error ellipse cross-section, the incompleteness 
introduced by this term alone is about 10$\%$ assuming Gaussian distribution. And it also may be affected by several 
factors:

1. We only select the SDSS targets with high confidence redshift (zConf $>$ 0.9) as our candidates, which will lead the 
probability that the targets without high quality redshift estimates will be rejected. The incompleteness increases 
from 1$\%$ for the bright objects to 6$\%$ in the faint end. 
  
2. Because of the target magnitude limit of the SDSS spectroscopic survey (Petrosian mag$\_$r $\leq$ 17.77 for 
main galaxies and PSF mag$\_$i $\leq$ 19.1 for quasars), there are also some optically faint LIGs which could not be 
included in the SDSS spectroscopic survey. So they are missed mainly due to their relatively higher redshift or serious 
obscuration by dust. 

3. There are also missing galaxies due to lack of fibers in dense regions, spectroscopic failures, and fiber collisions, 
which can be defined as the sampling rate: \~{f}$_{\rm t}$ $\sim$ 0.92 in average. (Blanton et al. 2001)

\section{"Likelihood Ratio" Method}
\label{sect:LR}
~~~~It is not easy to determine whether the matched SDSS targets are really the infrared objects or not. 
So we use the "Likelihood Ratio" (LR) method (Sutherland $\&$ Saunders 1992) to calculate the probability of 
the "true" cross-correlation for each matched SDSS object.
  
The likelihood ratio method is defined as that the cross-correlation probability between two observed sources 
are (assume that the errors are Gaussian in common)$\footnote{Note that the cross-scan errors for faint galaxies of 
{\it IRAS} are less Gaussian ({\it IRAS} Explanatory Supplement VII. Analysis of Processing C. Positional Accuracy), 
but this doesn't affect much on our statistical results of this work. So we use the Gaussian assumption and the 
Likelihood ratio method here and will try to improve it in further works.}$:
\begin{equation}
LR = \frac{Q(\leq m_{\rm i})exp(-r^{2}/2)}{2\pi\sigma_{\rm a}\sigma_{\rm b}n(\leq m_{\rm i})}
\label{eq:LRorigin}
\end{equation}
In this formula, r is the "normalized distance": 
\begin{equation}
r^{2} = \frac{(a1-a2)^{2}}{\sigma_{\rm a1}^{2}+\sigma_{\rm a2}^{2}} + \frac{(b1-b2)^{2}}{\sigma_{\rm b1}^{2}+\sigma_{\rm b2}^{2}} 
\label{eq:r}
\end{equation} 
(a1, b1) and (a2, b2) are the positions of each source, $\sigma$ terms standard deviations and n($\leq$ m$_{\rm i}$) 
is the local surface density of objects (galaxies) brighter than the candidate. The Q($\leq$ m$_{\rm i}$) is the 
multiplicative factor in the numerator which represents a priori probability that a "true" optical counterpart 
brighter than the flux limit exists amongst the identifications, and for simplicity we set Q = 1 in this work.

For our sample, the SDSS position uncertainties can be neglected compared with the {\it IRAS}'s large error ellipse. 
In this work, we refer to the {\it IRAS} uncertainty ellipse major axis (UncMaj) as $\sigma$$_{\rm a}$ , minor axis (UncMin) 
as $\sigma$$_{\rm b}$ and the position of the SDSS object in the {\it IRAS} 2$\sigma$ error ellipse (in the unit of 
$\sigma$, from 0 to 2) as r. We use the SDSS photometric targets to get n($\leq$ m$_{\rm i}$):
\begin{equation}
n(\leq m_{\rm i}) = \frac{N(\leq m_{\rm i})}{4\pi \sigma_{\rm a}\sigma_{\rm b}} 
\label{eq:N}
\end{equation} 
N($\leq$ m$_{\rm i}$) stands for the number of galaxies with r band magnitude less than or equal to the candidate's in the 
corresponding {\it IRAS} 2$\sigma$ error ellipse. Then we can get the LR formula for our sample:
\begin{equation}
LR = \frac{2exp(-r^{2}/2)}{N(\leq m_{\rm i})}
\label{eq:LRour}
\end{equation}

We calculate all of our samples' likelihood ratio values by using the SDSS photometric data (r band Petrosian magnitude 
for galaxies and i band PSF magnitude for QSOs). Then a random sample is selected for estimating the reliability of 
each object (use the method developed by Lonsdale et al. 1998; Rutledge et al. 2000; Masci et al. 2001), which is used 
to assess the cross-correlation probability and select a more reliable subsample. We also calculate the LRs and 
reliabilities for the PSCz sample 
(Saunders et al. 2000, all these optical targets selected from the PSC are identified as "true" IR objects) overlapped 
with our PSC sample for a comparison. The reliability distributions of the FSC, PSC and PSCz sample are shown in Fig. 2. 
%
   \begin{figure}
   \centering
   \includegraphics[width=\textwidth,height=100mm]{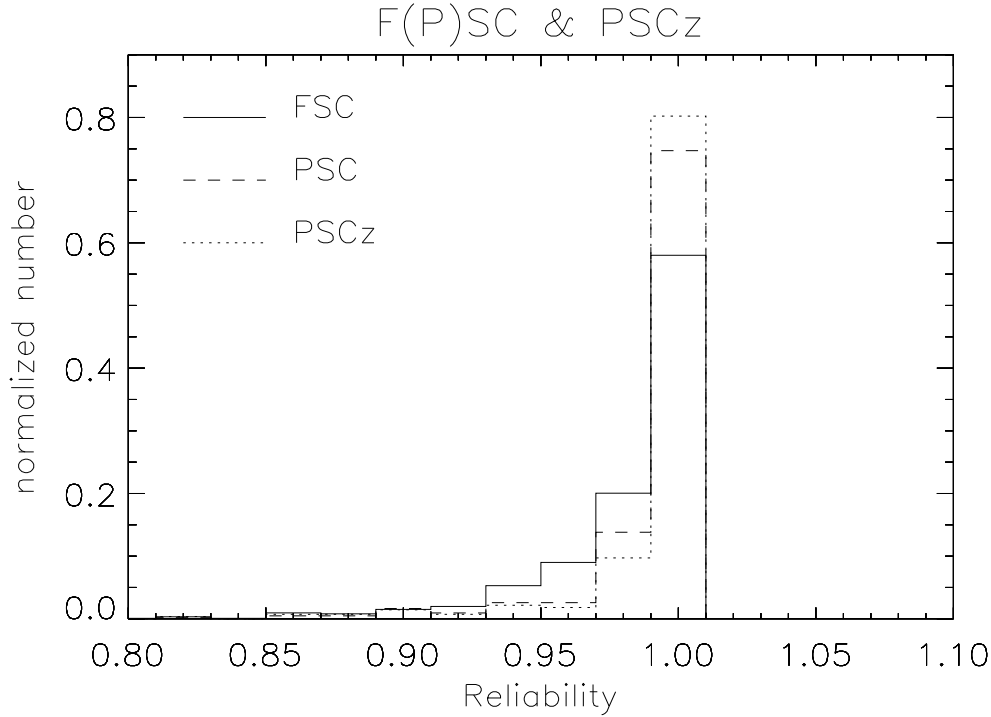}
   \caption{This figure shows the distributions of our sample's reliabilities. The solid line is the FSC sample's, 
the dashed line is the PSC sample's, and the dotted line is the PSCz sample's.}
   \label{Fig:plot2}
   \end{figure}

\section{The Catalog}
\label{sect:Catalog}
~~~~We present a Catalog (in ascii table) for the our sample of the LIGs, which contains the {\it IRAS}, 
SDSS-DR2 and FIRST informations. The structure and content of our Catalog $\footnote{The Catalog will be put on 
the web, use this URL:}$ are as follows:

The {\it IRAS} data (f(p)sciras.cat): the {\it IRAS} (FSC and PSC) name; {\it IRAS} RA and DEC; the error ellipse 
major (UncMaj), minor axis (UncMin)$\footnote{Note that the UncMaj and UncMin in the PSC stand for 1.96$\sigma$ 
significance level.}$ and position angle; 12, 25, 60 and 100 $\mu$m flux densities and qualities; and the calculated 
infrared luminosity using the SDSS spectral redshift.

The SDSS-DR2 photometric data (f(p)scsdssphoto1(2).cat): the SDSS ObjID; Photometric RA and DEC; objType and probPSF; 
SDSS five bands modelMag, psfMag, fiberMag, petroMag and their errors; Galactic extinctions; petroR50 and petroR90 
for band r.

The SDSS-DR2 spectroscopic data (f(p)scsdssspec1.cat): the SDSS SpecObjID; Spectroscopic RA and DEC; spectral redshift 
and its error; eclass and eCoeff; zWarning and zStatus; SpecClass, mjd, plate, fiberID.
 
The SDSS-DR2 emission line data (f(p)scsdssspec2.cat, from MPA-SDSS: www.mpa-garching.mpg.de/SDSS, Tremonti et al. 2004, 
Version 5.0$\_$4): the H$\alpha$, H$\beta$, [OII]$\lambda$$\lambda$3727,3729, 
[OIII]$\lambda$5007, [NII]$\lambda$6584, [SII]$\lambda$$\lambda$6716,6731 and [OI]$\lambda$6300 emission line's 
fluxes and flux errors; the corresponding Equivalent Widths (EQWs) and errors. Based on these data, we classify our 
sample into several spectral types: a) The galaxies without apparent emission lines (NoE for short) are chosen by 
the criterion: H$\alpha$ EQW $>$ -5$\AA$.$\footnote{Absorption lines have a positive sign.}$; b) The QSOs/Seyfert 
1s (S1) are those with Broad Line Regions (BLRs) and are also classified as QSOs by SDSS pipeline (specClass = 3); 
c) The classification of narrow emission line galaxies (Seyfert 2s, LINERs and HII regions) are performed using 
the emission line fluxes ratios, methods and the considered line ratios are: [OIII]$\lambda$5007/H$\beta$, 
[NII]$\lambda$6584/H$\alpha$, [SII]($\lambda$6716+$\lambda$6731)/H$\alpha$, [OI]$\lambda$6300/H$\alpha$ 
(Osterbrock 1985,1989; Wu et al. 1998b; Kauffmann et al. 2003c; Kewley et al. 2001). Specifically, for Seyfert 2s (S2): 
[OIII]/H$\beta$ $\geq$ 3; For LINERs (L): [NII]/H$\alpha$ $>$ 0.6, [SII]/H$\alpha$ $>$ 0.4, 
[OI]/H$\alpha$ $>$ 0.05 and [OIII]/H$\beta$ $<$ 3; For HII galaxies (H): [NII]/H$\alpha$ $<$ 0.6, [SII]/H$\alpha$ $<$ 
0.4, [OI]/H$\alpha$ $<$ 0.05 and [OIII]/H$\beta$ $<$ 3; The mixture types (LH: Mixture of LINERs and HIIs) are 
those which locate at the border of different spectral populations. The mixture type galaxies could be a transitional 
phase from HII galaxies to AGNs (Wu et al. 1998b). And there are also some galaxies which are not in the MPA's emission 
line catalog, so we classify them as Unknown (?). We will discuss this classification in detail in Sect. 5.3. 
 
The VLA FIRST radio data (f(p)scfirst.cat):  The VLA FIRST data (described in Sect. 2.4) contains: 
the FIRST name; FIRST RA and DEC; peak and integrated flux densities at 
1.4GHz; the local noise estimate; major and minor axis (FWHM), position angle; fitted MajAxis, MinAxis and PA before 
deconvolution; name of the coadded image containing the source; and based on the cross-correlation we give a "flags" 
for our sample: 0 stands for the case that the SDSS object is correlated with a FIRST source within 2" and 1 stands 
for that there are no FIRST counterparts in the corresponding search radius. 

We give each source a new index number for each FSC and PSC sample, and will do further works based on it. 

The main catalog (f(p)sc$\_$main.cat) contains only the most important informations we need, 
includes: the source number, the likelihood ratio (LR) and the Reliability we calculated in Sect.3, 
the {\it IRAS} name, the infrared luminosity, redshift, SpecObjID, Spectroscopic RA and DEC, SpecClass, ObjID, 
modelMag$\_$r, extinction$\_$r, petroMag$\_$r, the FIRST flag, the SDSS object's position in the {\it IRAS} error ellipse 
(in the unit of $\sigma$), the spectral types and the sign of the same sources across the two (FSC and PSC) sample.

\section{Analyses and Results}
\label{sect:analyses}
\subsection{Subsample Selection}
~~~~For the purpose of high confidence analyses we need a subsample with relatively high reliabilities for further works. 
From the comparison between our sample and the random sample (discussed in Sect.3 and shown in Fig. 2), here 
we give a selective criterion as the Reliability $\geq$ 0.98 for a relatively high cross-correlation probability. 
We choose this criterion for the subsample selection, and it contains 908 objects for FSC and 356 for PSC. 
From the comparison of the two redshifts (derived from our PSC sample, PSC subsample and the PSCz sample) of the 
same {\it IRAS} source (Fig. 3), we find that our subsample (at least the PSC) is more reliable because 
the sources' redshifts are consistent through the two sample except for only two sources. 
We also estimate our subsample's completeness from the LR distribution of the PSCz sample and find that it is 
about 86.69$\%$ if use the same selective criterion. 
%
   \begin{figure}
   \plottwo{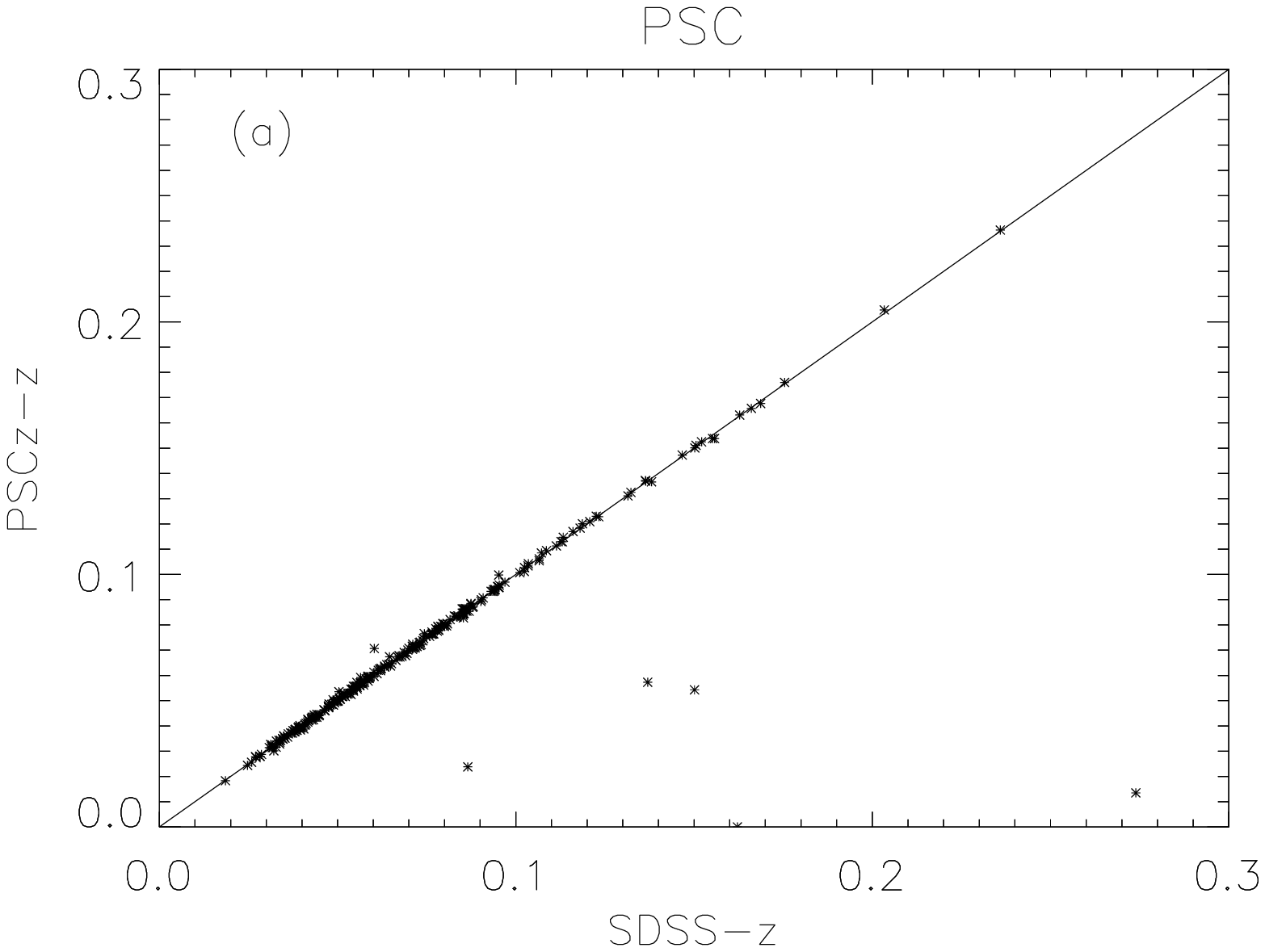} {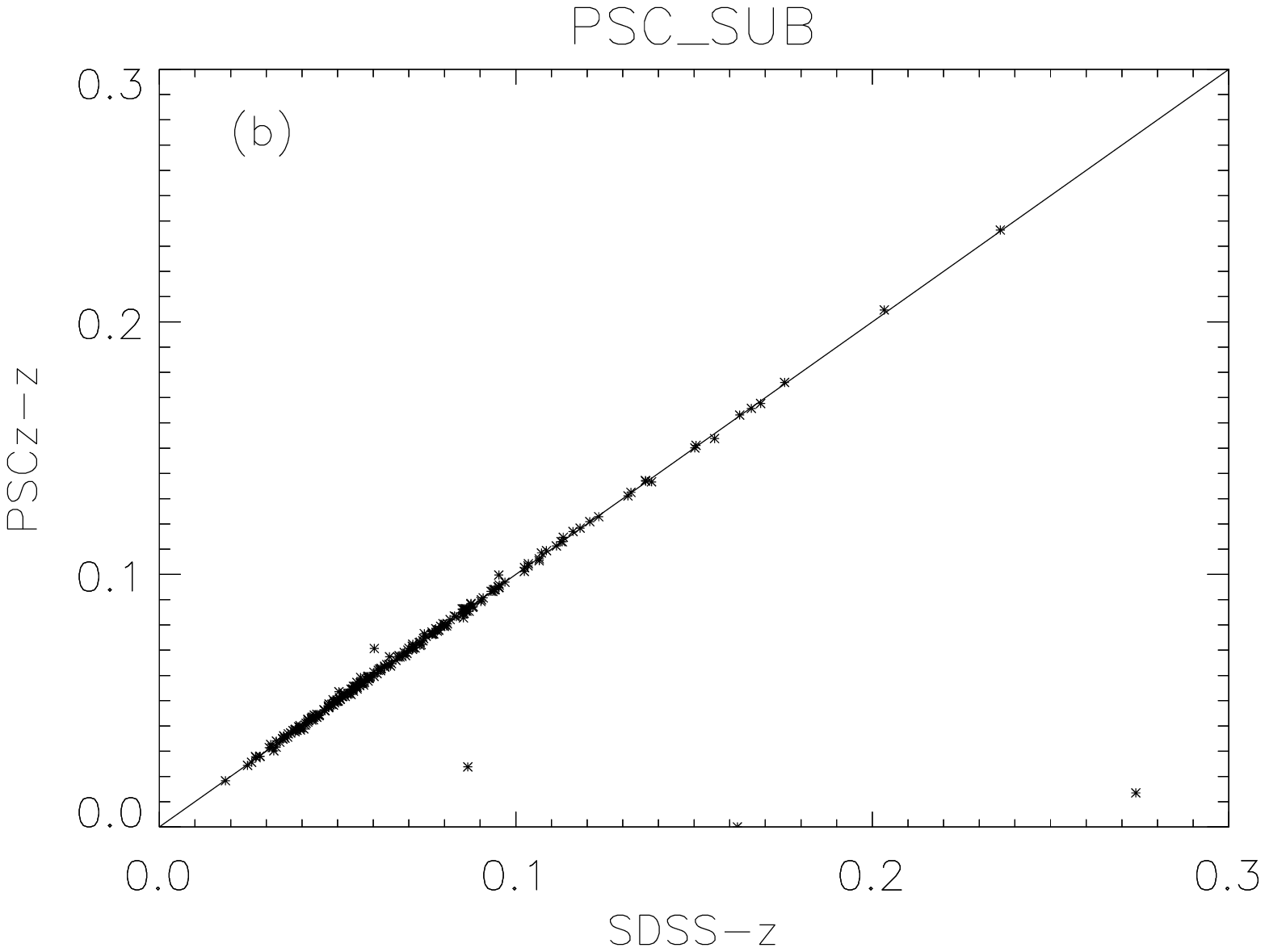}
   \caption{This figure shows the comparison between our PSC sample's and the PSCz sample's redshifts. (a):The PSC 
whole sample vs. PSCz; (b): The PSC subsample vs. PSCz.}
   \label{Fig:plot3}
\end{figure}
 
\subsection{Basic Statistical Properties}
~~~~The redshift and the L$_{\rm IR}$ distribution of our subsample are shown in Fig. 4 and Fig. 5. The number of LIGs 
(N$_{\rm LIGs}$, which L$_{\rm IR}$ $\sim$ 10$^{11}$ -- 10$^{12}$ L$_{\rm \odot}$) is 873 for FSC and 334 for PSC, and 
z$_{\rm median}$ $\sim$ 0.08 (FSC) and 0.05 (PSC). For the ULIGs (which L$_{\rm IR}$ $>$ 10$^{12}$ L$_{\rm \odot}$), N$_{\rm ULIGs}$ 
is 35 (FSC) and 22 (PSC), and z$_{\rm median}$ $\sim$ 0.18 (FSC) and 0.17 (PSC), $\sim$ 0.1 higher than the LIGs. The 
ratio N$_{\rm ULIGs}$:N$_{\rm LIGs}$ is 0.04 for FSC and a higher value 0.07 for PSC. For a comparison of the infrared 
luminosities derived from FSC and PSC (see Fig. 6), we find that the L$_{\rm IR}$ derived from FSC is consist with that 
from PSC by using the formula given in Sect 2.3.

The color (u-r) distributions of our subsample are shown in Fig. 7. Compared with the color separation of galaxy 
types described by Strateva et al. (2001), our result shows higher u-r values. The serious dust extinction of the 
LIGs, especially the ULIGs may be responsible for the redder color of our subsample.     
%
   \begin{figure}
   \plottwo{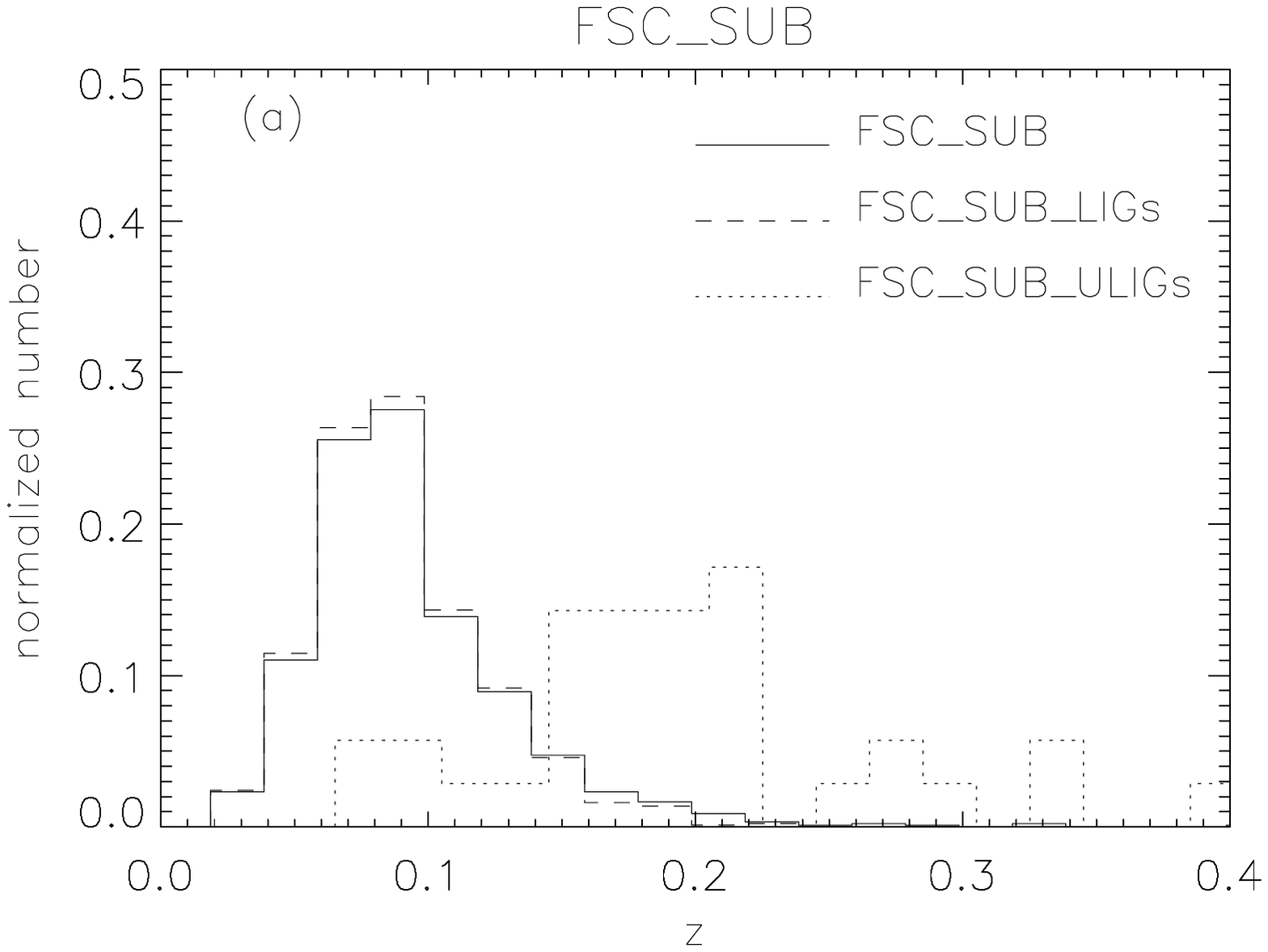} {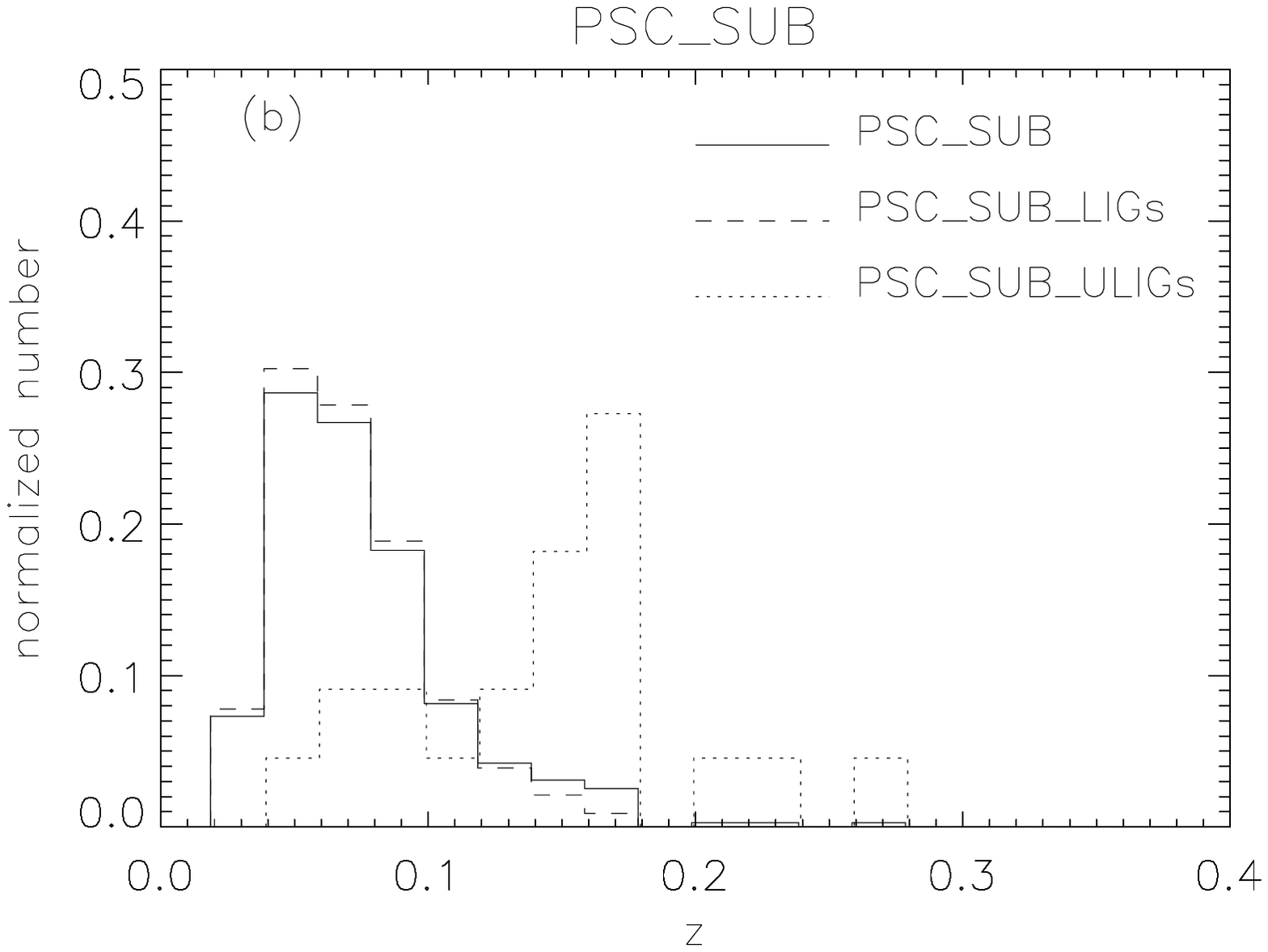}
   \caption{This figure shows the redshift distribution of our LIGs subsample. (a): The FSC subsample; 
(b): The PSC subsample. The solid lines are for the whole subsample, the dashed lines are for the 
LIGs (L$_{\rm IR}$ $\sim$ 10$^{11}$ -- 10$^{12}$ L$_{\rm \odot}$), and the dotted lines are for the 
ULIGs (L$_{\rm IR}$ $>$ 10$^{12}$ L$_{\rm \odot}$).}
   \label{Fig:plot4}
\end{figure}

%
\begin{figure}[h]
  \begin{minipage}[t]{0.5\linewidth}
  \centering
  \includegraphics[width=65mm,height=50mm]{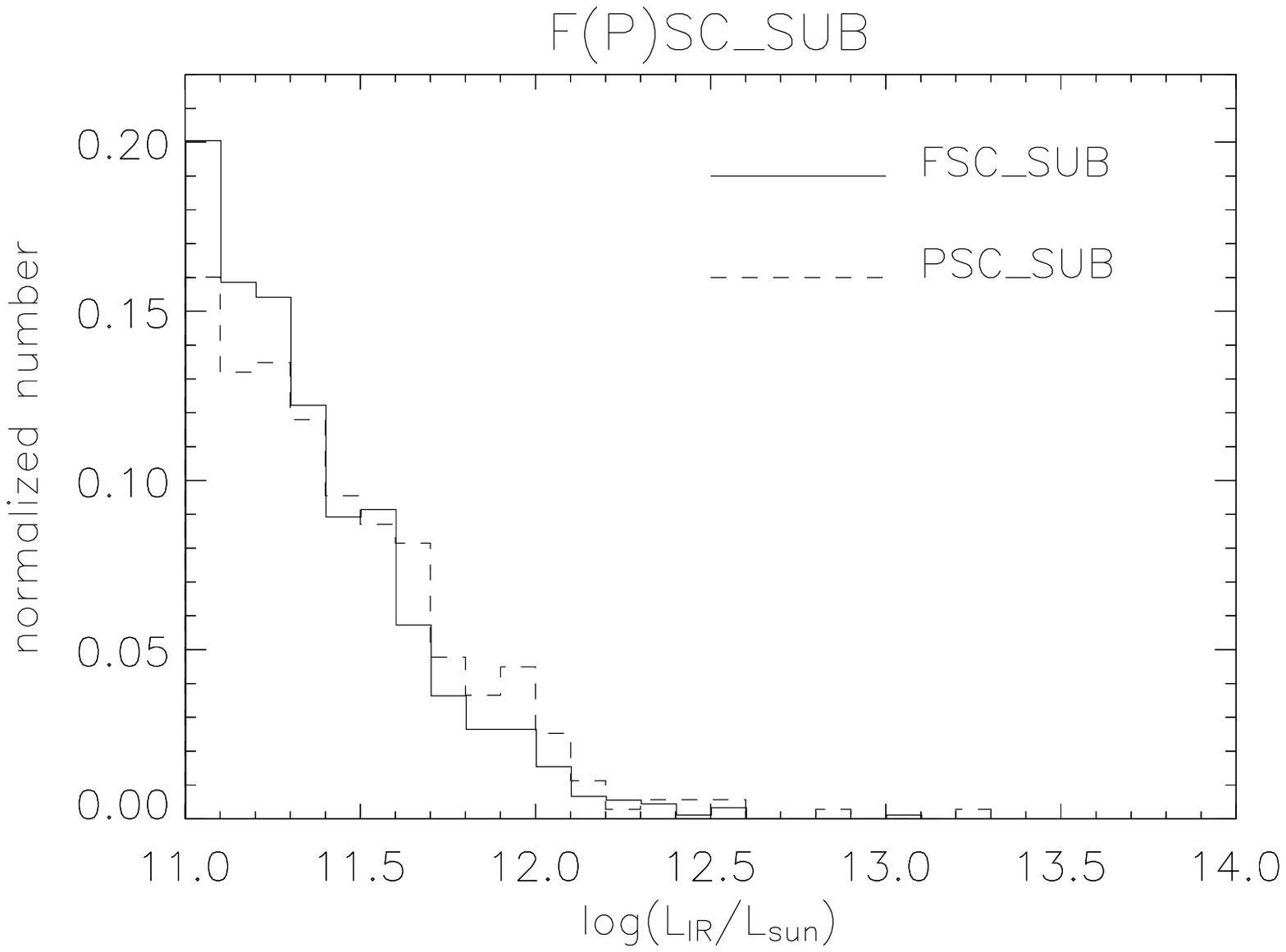}
  \vspace{-5mm}
  \caption{{\small This figure shows the infrared luminosity distribution of our LIGs subsample. The solid line is for 
the FSC, and the dashed line is for the PSC.} }
  \end{minipage}%
  \begin{minipage}[t]{0.5\textwidth}
  \centering
  \includegraphics[width=65mm,height=50mm]{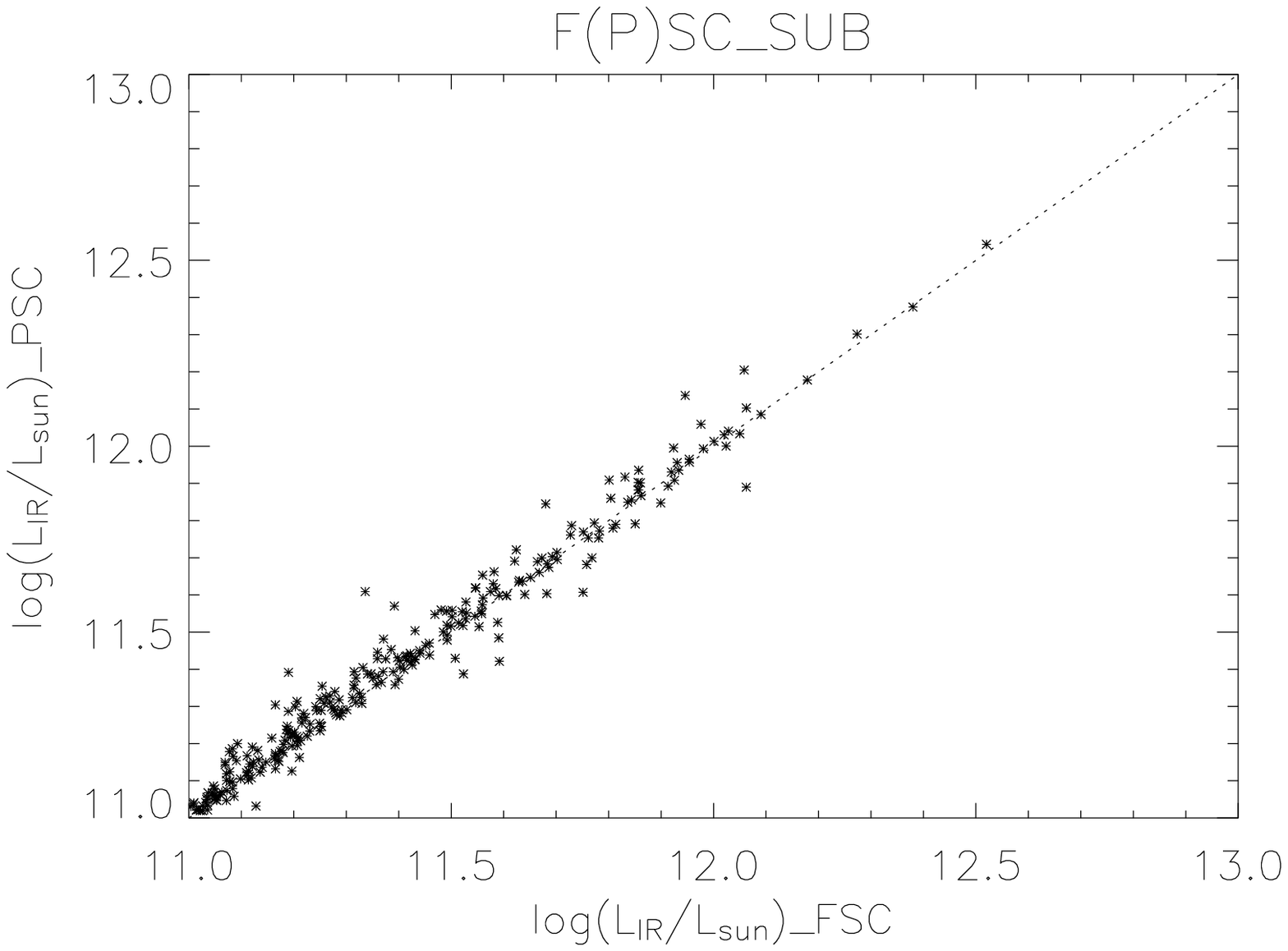}
  \vspace{-5mm}
  \caption{{\small This figure shows a comparison between the infrared luminosities derived from the FSC and the PSC 
subsample, all with 60$\mu$m flux qualities = 2 or 3.}}
  \end{minipage}%
  \label{Fig:fig5&6}
\end{figure}

%
   \begin{figure}
   \plottwo{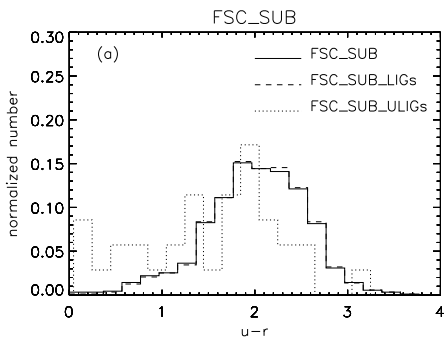} {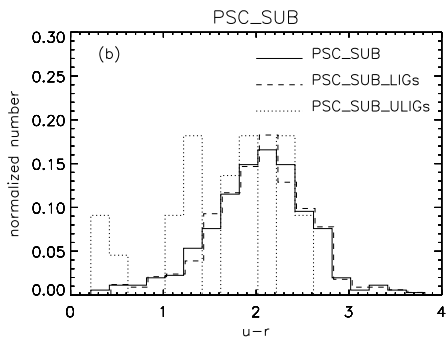}
   \caption{This figure shows the color (u-r) distribution of our LIGs subsample. (a): The FSC subsample; 
(b): The PSC subsample. The solid lines are for the LIGs (L$_{\rm IR}$ $\sim$ 10$^{11}$ -- 10$^{12}$ L$_{\rm \odot}$), 
and the dotted lines are for the ULIGs (L$_{\rm IR}$ $>$ 10$^{12}$ L$_{\rm \odot}$).}
   \label{Fig:plot6}
\end{figure}

\subsection{AGN Fraction}
~~~~Throughout this paper, we term AGNs as the assembly of the Seyfert 1s, Seyfert 2s, LINERs, and the Mixture types 
(S1+S2+L+LH, the spectral types are described in Sect.4). The BPT (Baldwin et al. 1981) diagrams for classifying 
the narrow emission line galaxies (Seyfert 2s (S2), LINERs (L), HIIs (H) and the Mixture types) are shown in Fig. 8. 
The number and fractions of each type are listed in Tables 1,2 and the distribution versus L$_{\rm IR}$ of our subsample 
is shown in Fig. 9 (the galaxies classified as Unknown(?) have been removed). Note that we have performed a volume 
correction by giving each objects a weight equal to the inverse of its maximum visibility volume: 1/Vmax (Schmidt 1968; 
Kauffmann et al. 2003ab), with a magnitude and flux cutoff for correcting the selection biases. 
We calculate the Vmax as follows: 
\begin{equation}
log D_{\rm l}(max)_{\rm SDSS} = \frac{mag_{\rm lim}-mag}{5}+log D_{\rm l}(z)
\label{eq:vmax1}
\end{equation}
\begin{equation}
D_{\rm l}(max)_{\rm IRAS} = D_{\rm l}(z)(\frac{f60}{f60_{\rm lim}})^{1/2}
\label{eq:vmax2}
\end{equation}
In this equation mag$_{\rm lim}$ is the SDSS magnitude cutoff (Petrosian mag$\_$r = 17.5), and f60$_{\rm lim}$ is the {\it IRAS} 
60$\mu$m flux cutoff (0.3Jy for FSC, 0.6Jy for PSC). Then the D$_{\rm l}$(max) for our estimation is the minimum of 
D$_{\rm l}$(max)$_{\rm SDSS}$ and D$_{\rm l}$(max)$_{\rm IRAS}$, so: Vmax = 4/3$\pi$D$_{\rm l}$$^{3}$(max).

%
   \begin{figure}
   \centering
   \includegraphics[width=\textwidth,height=100mm]{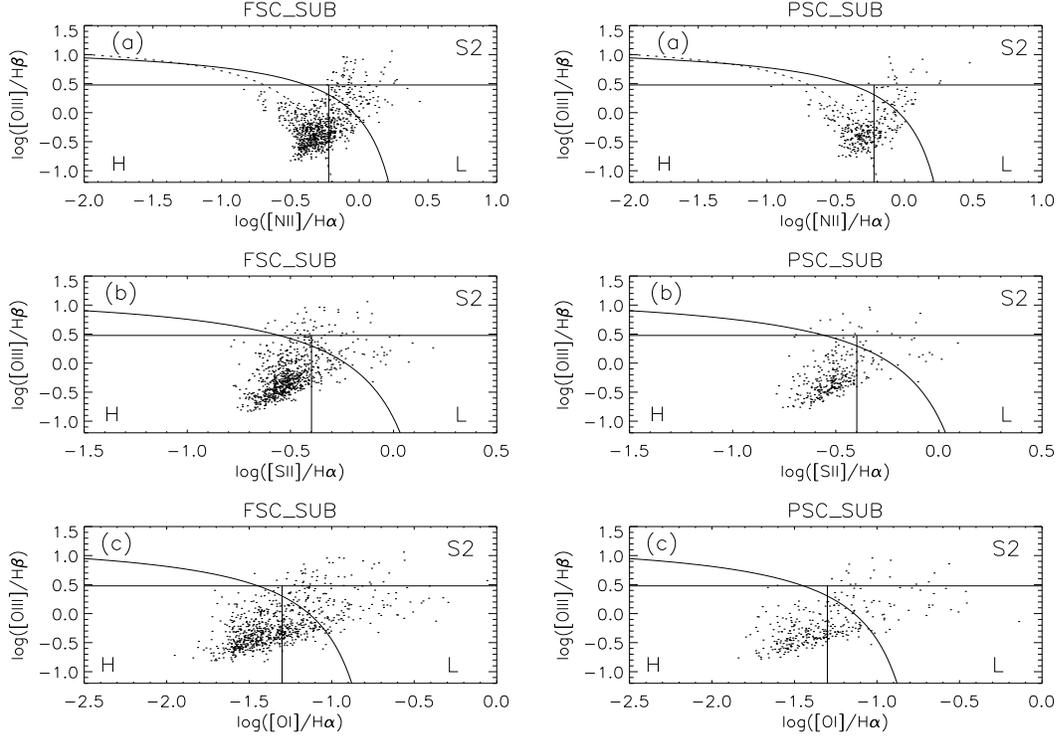}
   \caption{These figures show the BPT diagrams of our LIGs subsample. The straight lines are the criterion we used in 
this paper to classify the Seyfert 2s (S2) , LINERs (L) and HIIs (H). The solid curves are the criterion given by 
Kewley et al. 2001 for separating the starbursts and AGNs, and the dashed curves in figure (a) are the criterion 
given by Kauffmann et al. 2003c.}
   \label{Fig:plot8}
   \end{figure}

\begin{table}[]
  \caption[]{The spectral type distribution with the infrared luminosity of FSC subsample, the errors for AGN 
fractions are based on Poisson statistics.}
  \label{Tab:Spectype1}
  \begin{center}\begin{tabular}{c|cccc}
  \hline
Spectral Type & logL$_{\rm IR}$(L$_{\rm \odot}$) $\sim$ & 11.0-11.5 & 11.5-12.0 & $>$ 12.0   \\
  \hline
~S1$^{\mathrm{a}}$    & ~ & ~0.94$\%$(6)$^{\mathrm{b}}$ & 0.75$\%$(2) & 7.87$\%$(4)        \\
S2    & ~ & 5.22$\%$(22) & 3.89$\%$(5) & 0.00$\%$(0)        \\
L     & ~ & 9.05$\%$(27) & 8.41$\%$(10) & 5.53$\%$(1)      \\
LH    & ~ & 30.49$\%$(100) & 36.52$\%$(44) & 64.71$\%$(5)   \\
H     & ~ & 49.98$\%$(183) & 47.70$\%$(58) & 21.88$\%$(1)    \\
NoE   & ~ & 4.33$\%$(13) & 2.73$\%$(5) & 0.00$\%$(0)        \\
\hline
Total & ~ & 351 & 124 & 11  \\
\hline
AGN   & ~ & 45.69$\pm$3.67$\%$(155) & 49.57$\pm$6.35$\%$(61) & 78.12$\pm$24.70$\%$(10)  \\
\hline
  \end{tabular}\end{center}
   \begin{list}{}{}
   \item[$^{\mathrm{a}}$]  The spectral types: S1, S2, L, LH, H and NoE stand for the Seyfert 1s, Seyfert 2s, LINERs, 
Mixture types and HIIs as described in Sect.4.
   \item[$^{\mathrm{b}}$]  The volume corrected fraction of different spectral types in each L$_{\rm IR}$ bin, the 
number of each type galaxies is in the bracket.
   \end{list}
\end{table}
\begin{table}[]
  \caption[]{The spectral type distribution with the infrared luminosity of PSC subsample.}
  \label{Tab:Spectype2}
  \begin{center}\begin{tabular}{c|cccc}
  \hline
Spectral Type & logL$_{\rm IR}$(L$_{\rm \odot}$) $\sim$ & 11.0-11.5 & 11.5-12.0 & $>$ 12.0   \\
  \hline
S1    & ~ & 1.59$\%$(3) & 2.59$\%$(2) & 3.69$\%$(3)        \\
S2    & ~ & 6.31$\%$(7) & 6.43$\%$(3) & 0.00$\%$(0)        \\
L     & ~ & 7.34$\%$(9) & 4.74$\%$(2) & 0.00$\%$(0)      \\
LH    & ~ & 37.53$\%$(47) & 33.83$\%$(22) & 74.94$\%$(6)        \\
H     & ~ & 45.53$\%$(70) & 50.06$\%$(30) & 21.37$\%$(1)     \\
NoE   & ~ & 1.69$\%$(4) & 2.34$\%$(2) & 0.00$\%$(0)        \\
\hline
Total & ~ & 140 & 61 & 10  \\
\hline
AGN   & ~ & 52.78$\pm$6.50$\%$(66) & 47.60$\pm$8.84$\%$(29) & 78.63$\pm$26.21$\%$(9)  \\
\hline
  \end{tabular}\end{center}
\end{table}
%
   \begin{figure}
   \plottwo{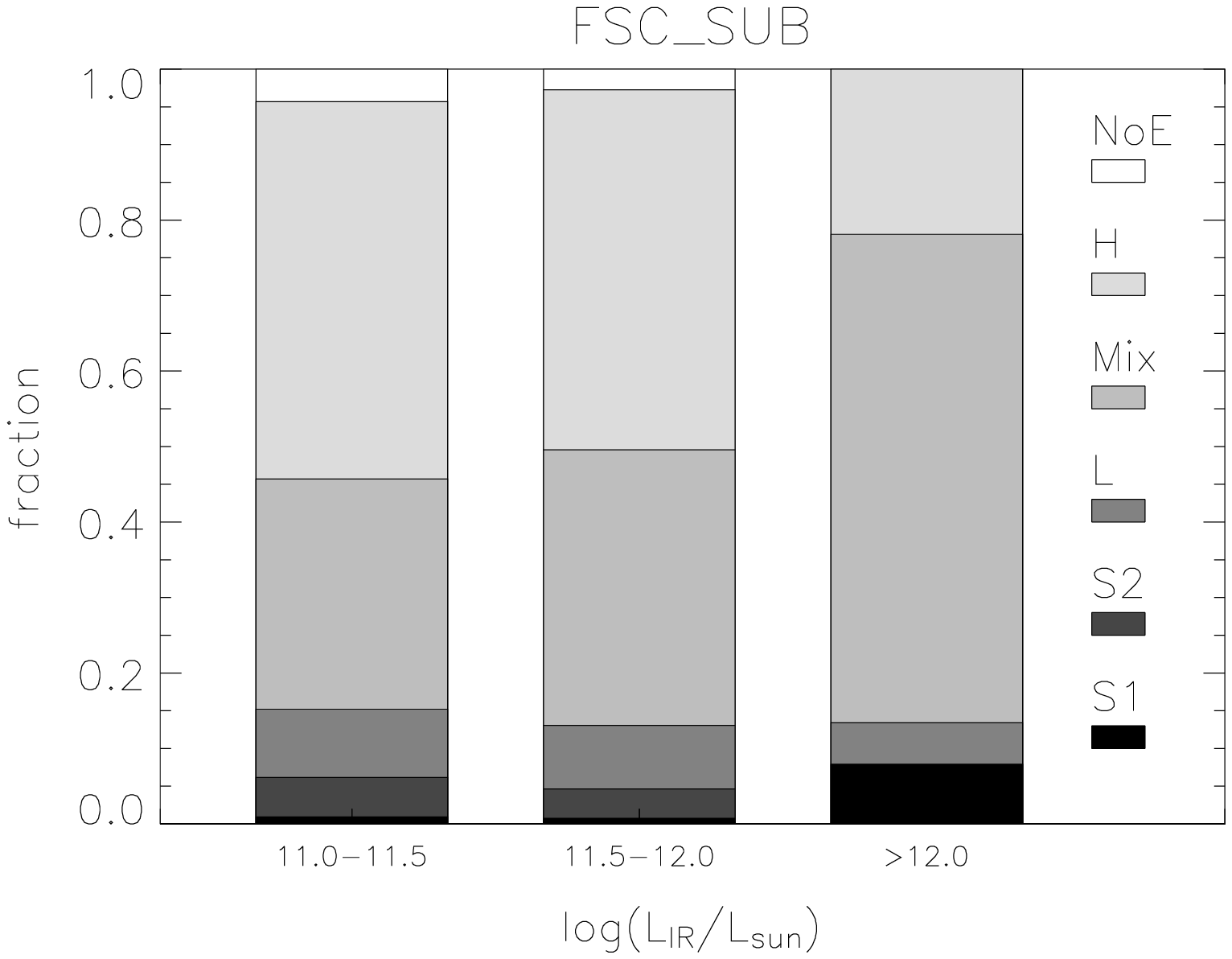} {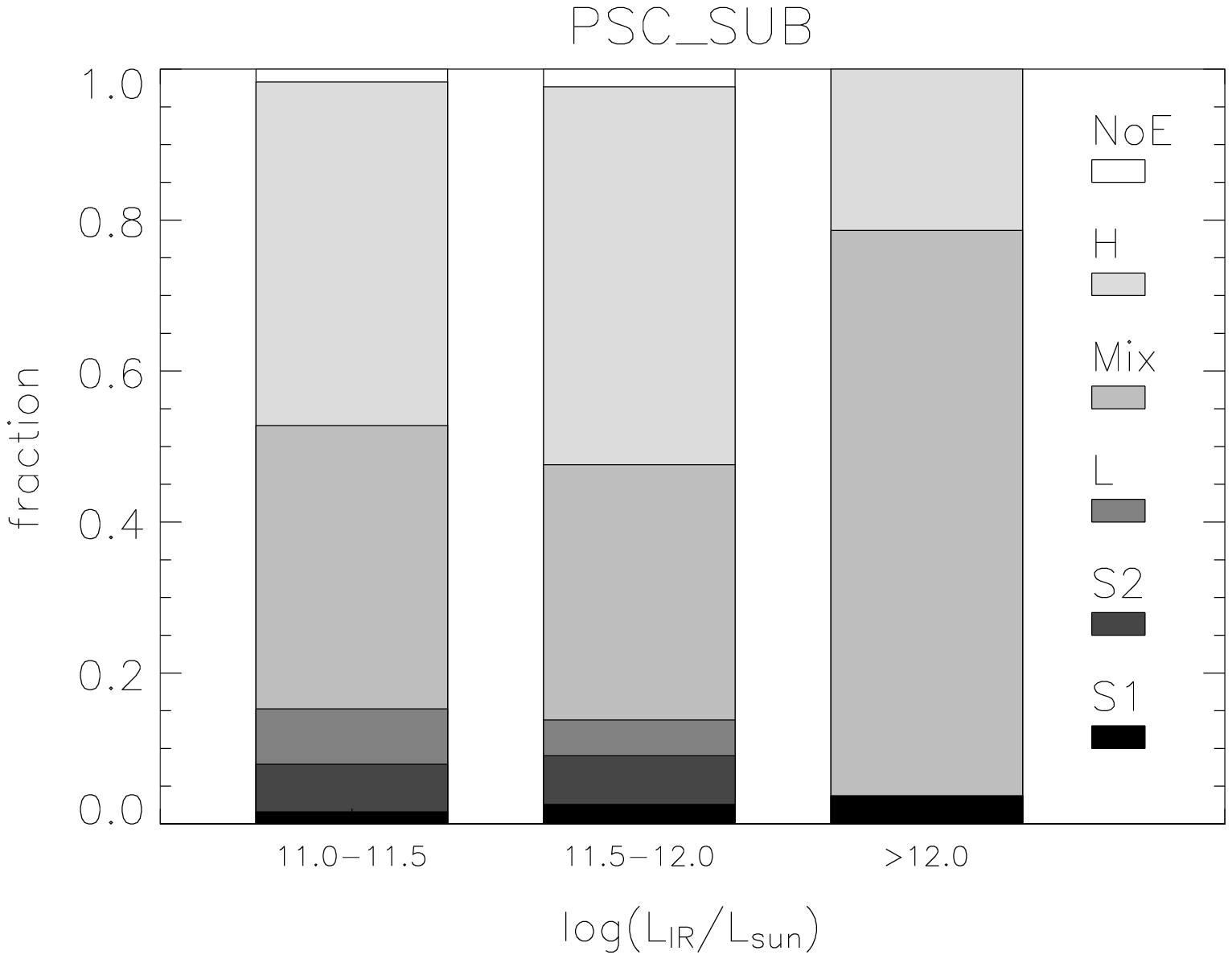}
   \caption{This figure shows the spectral type (Seyfert 1s (S1), Seyfert 2s (S2), LINERs (L), Mixture types (Mix), 
HIIs (H) and No apparent Emission lines (NoE), colored from black to white) distribution vs. the infrared luminosity 
of our subsample. Left: The FSC subsample; Right: The PSC subsample.}
   \label{Fig:plot9}
\end{figure}

The AGN fractions of our subsample increase with the infrared luminosities, from $\sim$45$\%$ to 80$\%$ 
when L$_{\rm IR}$ increases from 10$^{11}$ to 10$^{13}$ L$_{\rm \odot}$. This is in agreement with the previous results that 
the AGN fraction increases from the LIGs to ULIGs, from 47$\%$ to 70-75$\%$ (Kim et al. 1995; Veilleux et al. 
1995,1999) and 56$\%$ to 82$\%$ (Wu et al. 1998b). 
From Tables 3,4 we also find that some galaxies without apparent emission lines (NoE) have high L$_{\rm IR}$, especially 
for PSC subsample (due to their relative higher L$_{\rm IR}$). These galaxies may be either: a) Have low S/N ratios 
or bad spectra; b) One member of a galaxy pair or group, and the large amount of infrared emissions may come from its 
companions; c) Have late stage merger feature and e(a) spectral feature (Poggianti \& Wu 2000) or E+A feature, 
which indicates a post-starburst phase (Zabludoff et al. 1996; Yang et al. 2004; Goto 2005a).

\subsection{Infrared to Radio Correlation}
~~~~The infrared to radio correlation of our subsample is shown in Fig. 10, we calculate the L$_{\rm 60 \mu m}$ and 
L$_{\rm 1.4GHz}$ using the formula (Yun et al. 2001):
\begin{equation}
log L_{\rm 60\mu m}(L_{\rm \odot}) = 6.014 + 2log D + log S_{\rm 60\mu m}  
\label{eq:11}
\end{equation} 
\begin{equation}
log L_{\rm 1.4GHz}(WHz^{-1}) = 20.08 + 2log D + log S_{\rm 1.4GHz}  
\label{eq:12}
\end{equation}
where D is the luminosity distance in Mpc and S$_{\rm 60 \mu m}$ and S$_{\rm 1.4GHz}$ are flux densities in units of Jy. The 
straight line is the best fitting line obtained by Yun et al. (2001) for an all-sky sample of infrared detected 
galaxies from {\it IRAS}:
\begin{equation}
log L_{\rm 1.4GHz} = (0.99 \pm 0.01) log (L_{\rm 60\mu m}/L_{\rm \odot}) + (12.07 \pm 0.08) 
\label{eq:13} 
\end{equation}
From these relations we find that the infrared to radio correlation for our subsample are follow the correlations 
for an all-sky sample of infrared detected galaxies from {\it IRAS} (Yun et al 2001). The slight deviation for the 
PSC\_SUB is not significant and smaller than the scattering of the infrared to radio correlation.

The q parameter is also plotted for our subsample in Fig. 11, following the formula (Condon et al. 1991):
\begin{equation}
q = log (\frac{2.58S_{\rm 60\mu m} + S_{\rm 100\mu m}}{2.98Jy}) - log (\frac{S_{\rm 1.4GHz}}{Jy})
\label{eq:14}
\end{equation}
The solid line is at q = 2.34 which is the mean value obtained by Yun et al. (2001), the top and bottom dotted lines are 
limits for three times FIR excess and radio excess from the mean respectively. The radio excess objects are mainly 
Radio Loud (RL) AGNs (Roy \& Norris 1997) that may have some complex mechanisms of energy generation 
(e.g. the jet emission).
%
   \begin{figure}
   \plottwo{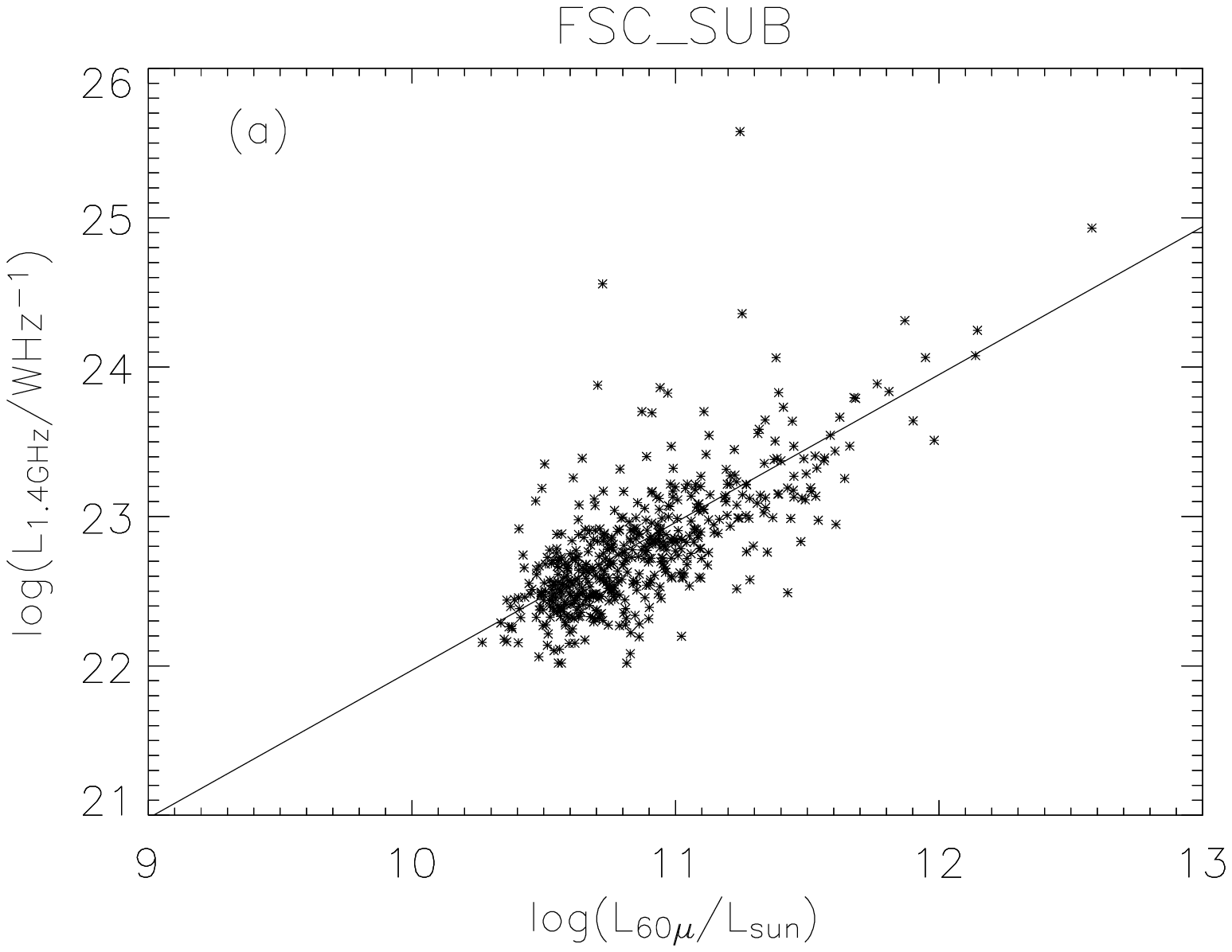} {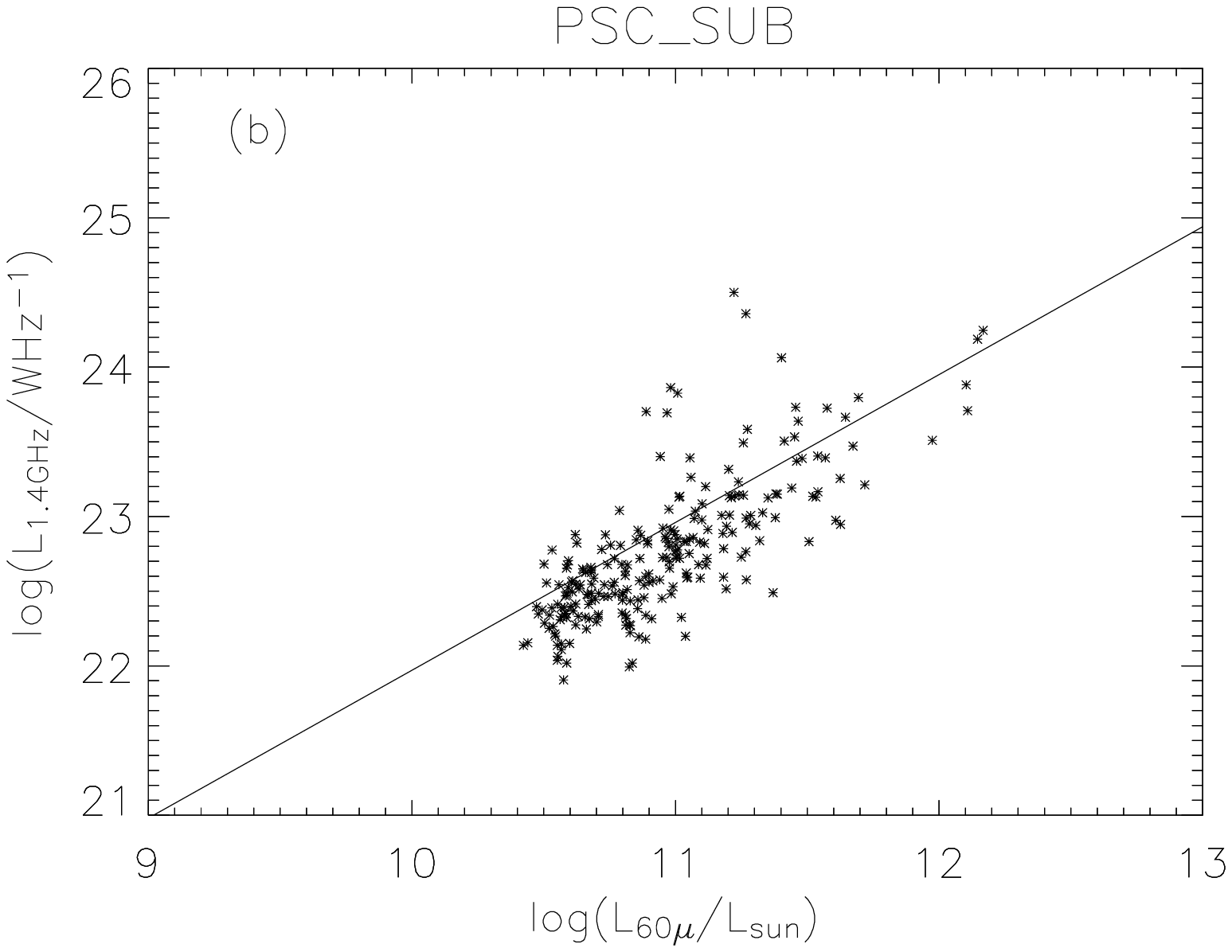}
   \caption{This figure shows the Infrared (60$\mu$m) to Radio (1.4GHz) correlation of the subsample. (a): The FSC 
subsample; (b): The PSC subsample. The straight line is the best fit obtained by Yun et al. (2001) for an all-sky sample 
of infrared detected galaxies from {\it IRAS}.}
   \label{Fig:plot9}
\end{figure}
%
   \begin{figure}
   \plottwo{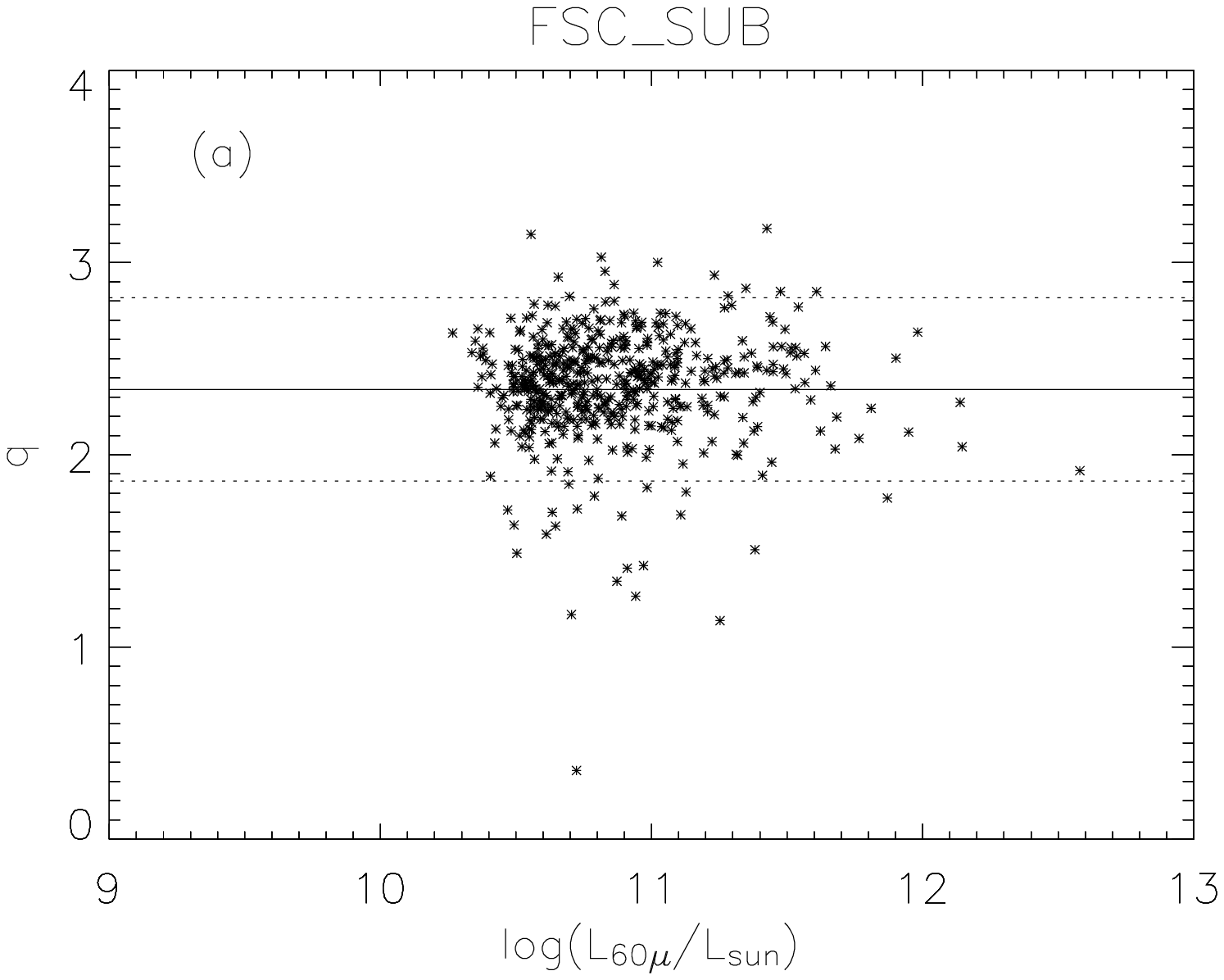} {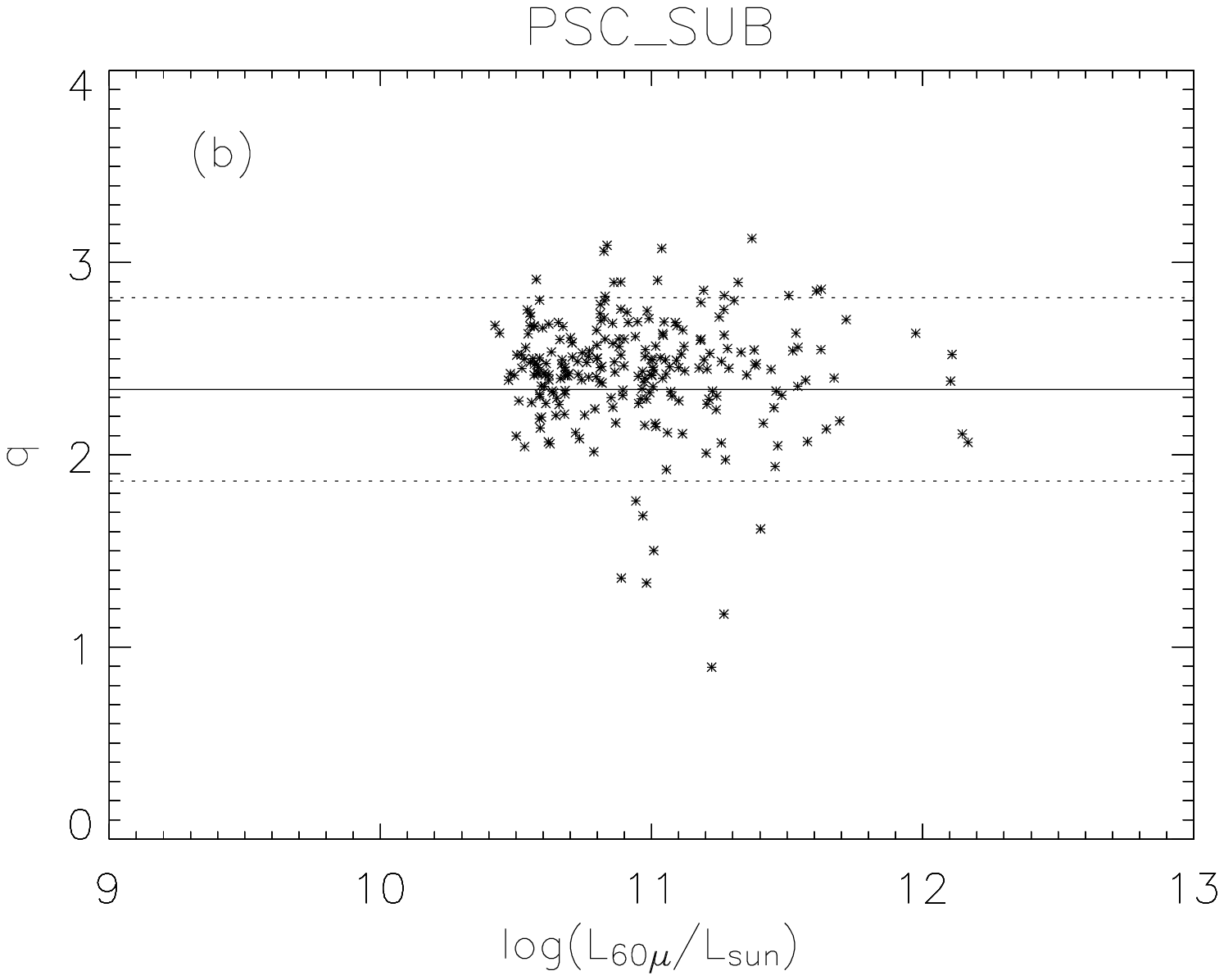}
   \caption{This figure shows the q parameter for the subsample. (a): The FSC subsample; (b): The PSC subsample. 
The solid line is at q = 2.34 which is the mean value obtained by Yun et al. (2001), the top and bottom dotted lines are 
limits for three times FIR excess and radio excess from the mean respectively.}
   \label{Fig:plot10}
\end{figure}

\section{Summary}
\label{sect:Summary}
~~~~In this paper we select a sample of Luminous Infrared Galaxies based on the cross-correlation between the 
{\it IRAS} FSC and PSC data and the SDSS-DR2, and present a Catalog. We use the "likelihood ratio" method 
to estimate the sample's reliability and for a high confidence subsample selection. 
Although the LR method also has some problems and needs to be improved, it seems that it can be used as a stable 
and creditable sample selection method based on the analyses and comparison in this work. From the statistical 
analyses (e.g., the redshift, L$_{\rm IR}$ and color distributions, the spectral types, and the radio to infrared 
correlations) we find that the LIGs and ULIGs are quite different. We will perform further analyses in the future 
and attempt to know more about the LIGs, such as their morphologies and environments (Wang et al. in preparation), 
the origins of the IR excess (Pasquali et al. 2005) and their star formation histories. Some interesting subsamples 
like the IR QSOs (Zheng et al. 2002; Hao et al. 2005) and RL AGNs (Best et al. 2005) will also be selected 
and analyzed for understanding the connections between the star formation and AGN activity. During such works 
we will keep on finding better statistical methods for huge astronomical data mining and analyses.

\begin{acknowledgements}
We would like to thank Drs. J. Y. Wei, S. Mao, S. Komossa, J. Wang for advice and helpful 
discussions. We also thank the anonymous referee and the editor for very constructive comments and suggestions. 
This project is supported by NSF of China No.10273012, No.10333060, No.10473013 and NKBRSF G1999075404. 
Funding for the creation and distribution of the SDSS Archive has been provided by the Alfred P. Sloan Foundation, 
the Participating Institutions, the National Aeronautics and Space Administration, the National Science Foundation, 
the U.S. Department of Energy, the Japanese Monbukagakusho, and the Max Planck Society. The SDSS Web site 
is http://www.sdss.org/. The SDSS is managed by the Astrophysical Research Consortium (ARC) for the Participating 
Institutions. The Participating Institutions are The University of Chicago, Fermilab, the Institute for Advanced 
Study, the Japan Participation Group, The Johns Hopkins University, the Korean Scientist Group, Los Alamos National 
Laboratory, the Max-Planck-Institute for Astronomy (MPIA), the Max-Planck-Institute for Astrophysics (MPA), New 
Mexico State University, University of Pittsburgh, University of Portsmouth, Princeton University, the United 
States Naval Observatory, and the University of Washington. This work also used the {\it IRAS} data from the PSC 
(IPAC 1986) and FSC (Moshir+ 1989), the FIRST data from the VLA-FIRST project and the emission line data from the 
SDSS studies at MPA/JHU. 

\end{acknowledgements}

\label{lastpage}


\begin{thebibliography}{99}
  
  \bibitem[2004]{Ab04} Abazajian K., Adelman-McCarthy J. K., Ag\"{u}eros M. A. et al., 2004, AJ, 128, 502

  \bibitem[1981]{Ba81} Baldwin J. A., Phillips M. M., Terlevich R., 1981, PASP, 93, 5 (BPT)

  \bibitem[1995]{Be95} Becker R. H., White R. L., Helfand D. J., 1995, ApJ, 450, 559

  \bibitem[2005]{Be05} Best P. N., Kauffmann G., Heckman T. M., Ivezi\'{c} \v{Z}., 2005, MNRAS, 362, 9

  \bibitem[2001]{Bl01} Blanton M. R., Dalcanton J., Eisenstein D. et al., 2001, AJ, 121, 2358

  \bibitem[2000]{Cal00} Calzetti D., Armus L., Bohlin R. C. et al., 2000, ApJ, 533, 682

  \bibitem[2000]{Col00} Cole S., Lacey C. G., Baugh C. M., Frenk C. S., 2000, MNRAS, 319, 168

  \bibitem[1991]{Con91} Condon J. J., Anderson M. L., Helou G., 1991, ApJ, 376, 95

  \bibitem[1992]{Con92} Condon J. J., 1992, ARA\&A, 30, 575

  \bibitem[2005]{Go05a} Goto T., 2005a, MNRAS, 357, 937

  \bibitem[2005]{Go05b} Goto T., 2005b, MNRAS, 360, 322

  \bibitem[2005]{Hao05} Hao C. N., Xia X. Y., Mao S., Wu H., Deng Z. G., 2005, ApJ, 625, 78

  \bibitem[1985]{He85} Helou G., Soifer B. T., Rowan-Robinson M., 1985, ApJ, 298, L7

  \bibitem[1988]{He88} Helou G., Khan I., Malek L., Boehmer L., 1988, ApJS, 68, 151

  \bibitem[1993]{He93} Helou G., Bicay M. D., 1993, ApJ, 415, 93

  \bibitem[2002]{Iv02} Ivezi\'{c} \v{Z}., Menou K., Knapp G. R. et al., 2002, AJ, 124, 2364

  \bibitem[2003]{Ka03a} Kauffmann G., Heckman T. M., White S. D. M. et al., 2003a, MNRAS, 341, 33

  \bibitem[2003]{Ka03b} Kauffmann G., Heckman T. M., White S. D. M. et al., 2003b, MNRAS, 341, 54

  \bibitem[2003]{Ka03c} Kauffmann G., Heckman T. M., Tremonti C. et al., 2003c, MNRAS, 346, 1055

  \bibitem[2001]{Ke01} Kewley L. J., Dopita M. A., Sutherland R. S., Heisler C. A., Trevena J., 2001, ApJ, 556, 121

  \bibitem[1995]{Kim95} Kim D. -C., Sanders D. B., Veilleux S., Mazzarella J. M., Soifer B. T., 1995, ApJS, 98, 129

  \bibitem[1998]{Kim98a} Kim D. -C., Sanders D. B., 1998a, ApJS, 119, 41

  \bibitem[1998]{Kim98b} Kim D. -C., Veilleux S., Sanders D. B., 1998b, ApJ, 508, 627

  \bibitem[2002]{Kim02} Kim D. -C., Veilleux S., Sanders D. B., 2002, ApJS, 143, 277

  \bibitem[1989]{La89} Lawrence A., Rowan-Robinson M., Leech K., Jones D. H., Wall J. V., 1989, MNRAS, 240, 329

  \bibitem[1998]{Lo98} Lonsdale C. J. et al., 1998, B. McLean, D. Golombek, J. Hayes, H. Payne, ed., IAU Colloq. 179, 
New Horizons from Multi-Wavelength Sky Surveys, Dordrecht: Kluwer, 450

  \bibitem[2001]{Ma01} Masci F. J., Condon J. J., Barlow T. A. et al., 2001, PASP, 113, 10

  \bibitem[1990]{Mo90} Moshir M., Kopan G., Conrow T. et al., 1990, IRAS Faint Source Catalogue, v. 2.0

  \bibitem[1984]{Ne84} Neugebauer G., Habing H. J., van Duinen R. et al., 1984, ApJ, 278, L1

  \bibitem[1989]{Os89} Osterbrock D. E., 1989, Astrophysical of Gaseous Nebulae and Active Galactic Nuclei, 
Mill Valley CA: University Science Books

  \bibitem[1985]{Os85} Osterbrock D. E., Pogge R. W., 1985, ApJ, 297, 166

  \bibitem[2005]{Pas05} Pasquali A., Kauffmann G., Heckman T. M., 2005, MNRAS, 361, 1121

  \bibitem[2000]{Po00} Poggianti B. M., Wu H., 2000, ApJ, 529, 157

  \bibitem[1997]{Roy97} Roy A. L., Norris R. P., 1997, MNRAS, 289, 824

  \bibitem[2000]{Ru00} Rutledge R. E., Brunner R. J., Prince T. A., Lonsdale C., 2000, ApJS, 131, 335

  \bibitem[1988]{San88} Sanders D. B., Soifer B. T., Elias J. H. et al., 1988, ApJ, 325, 74

  \bibitem[1996]{San96} Sanders D. B., Mirabel I. F., 1996, ARA\&A, 34, 749

  \bibitem[2000]{Sau00} Saunders W., Sutherland W. J., Maddox S. J. et al., 2000, MNRAS, 317, 55

  \bibitem[1968]{Sch68} Schmidt M., 1968, ApJ, 151, 393

  \bibitem[1987]{So87a} Soifer B. T., Houck J. R., Neugebauer G., 1987a, ARA\&A, 25, 187

  \bibitem[1987]{So87b} Soifer B. T., Sanders D. B., Madore B. F. et al., 1987b, ApJ, 320, 238

  \bibitem[2001]{Strat01} Strateva I., Ivezi\'{c} \v{Z}., Knapp G. R. et al., 2001, AJ, 122, 1861

  \bibitem[1992]{Su92} Sutherland W., Saunders W., 1992, MNRAS, 259, 413

  \bibitem[2004]{Tr04} Tremonti C. A., Heckman T. M., Kauffmann G. et al., 2004, ApJ, 613, 898

  \bibitem[1995]{Ve95} Veilleux S., Kim D. -C., Sanders D. B., Mazzarella J. M., Soifer B. T., 1995, ApJS, 98, 171

  \bibitem[1999]{Ve99} Veilleux S., Kim D. -C., Sanders D. B., 1999, ApJ, 522, 113

  \bibitem[1997]{Wh97} White R. L., Becker R. H., Helfand D. J., Gregg M. D., 1997, ApJ, 475, 479

  \bibitem[1998]{Wu98a} Wu H., Zou Z. L., Xia X. Y., Deng Z. G., 1998a, A\&AS, 127, 521

  \bibitem[1998]{Wu98b} Wu H., Zou Z. L., Xia X. Y., Deng Z. G., 1998b, A\&AS, 132, 181

  \bibitem[2004]{Ya04} Yang Y., Zabludoff A. I., Zaritsky D., Lauer T. R., Mihos J. C., 2004, ApJ, 607, 258

  \bibitem[2000]{Yo00} York D. G., Adelman J., Anderson J. E. Jr. et al., 2000, AJ, 120, 1579

  \bibitem[2001]{Yun01} Yun M. S., Reddy N. A., Condon J. J., 2001, ApJ, 554, 803

  \bibitem[1996]{Za96} Zabludoff A. I., Zaritsky D., Lin Huan et al., 1996, ApJ, 466, 104

  \bibitem[2002]{ZhXZ02} Zheng X. Z., Xia X. Y., Mao S., Wu H., Deng Z. G., 2002, AJ, 124, 18

  \bibitem[2004]{ZhXZ04} Zheng X. Z., Hammer F., Flores H., Ass\'{e}mat, F., Pelat, D., 2004, A\&A, 421, 847
  
  \bibitem[1991]{Zou91} Zou Z. L., Xia X. Y., Deng Z. G., Su H. J., 1991, MNRAS, 252, 593

\end{thebibliography}
\end{document}